\documentclass[10pt]{article}
\pdfoutput=1
\usepackage{amsmath,amssymb,amsfonts,amsthm,amsbsy}
\usepackage{fullpage}
\usepackage[usenames]{color}
\usepackage{ulem}
\usepackage{epsfig}
\usepackage{bm}
\usepackage{natbib}
\usepackage{microtype}

\newcommand{\be}{\begin{equation}} 
\newcommand{\ee}{\end{equation}}

\usepackage{hyperref}
\usepackage{cite}
\usepackage{xr} 
\usepackage{algpseudocode}
\usepackage{algorithmicx}
\usepackage{algorithm}

\title{A structured matrix factorization framework for large scale calcium imaging data analysis}

\author{Eftychios A. Pnevmatikakis \and Yuanjun Gao \and Daniel Soudry \and David Pfau \and Clay Lacefield \and Kira Poskanzer \and Randy Bruno \and Rafael Yuste \and Liam Paninski \\ \small Columbia University}
\date{}

\begin{document}

\maketitle

\begin{abstract}
We present a structured matrix factorization approach to analyzing calcium imaging recordings of large neuronal ensembles. Our goal is to simultaneously identify the locations of the neurons, demix spatially overlapping components, and denoise and deconvolve the spiking activity of each neuron from the slow dynamics of the calcium indicator. The matrix factorization approach relies on the observation that the spatiotemporal fluorescence activity can be expressed as a product of two matrices: a spatial matrix that encodes the location of each neuron in the optical field and a temporal matrix that characterizes the calcium concentration of each neuron over time. We present a simple approach for estimating the dynamics of the calcium indicator as well as the observation noise statistics from the observed data. These parameters are then used to set up the matrix factorization problem in a constrained form that requires no further parameter tuning. We discuss initialization and post-processing techniques that enhance the performance of our method, along with efficient and largely parallelizable algorithms. We apply our method to {\it in vivo} large scale multi-neuronal imaging data and also demonstrate how similar methods can be used for the analysis of {\it in vivo} dendritic imaging data.
\end{abstract}

\section{Introduction}
Calcium imaging is becoming a standard tool for monitoring large neuron populations. Over the recent years exciting developments have enabled whole brain imaging of small animals \citep{AHR13,prevedel2014} at reasonable imaging rates. On a different front, engineering of genetically encoded calcium indicators continues, offering increasingly sensitive indicators that can reliably detect single action potentials in {\it in vivo} conditions \citep{gcamp6}. These developments pose significant challenges from a statistical viewpoint. The data analyst typically faces three major problems: (i) identifying the region of interest (ROI) of each neuron in the optical field, (ii) demixing spatially overlapping ROIs (where overlap is due either to the projection of a 3$d$ volume onto a 2$d$ imaging plane, or to insufficient spatial resolution in 3$d$ imaging methods) and (iii) deconvolving (and denoising) the spiking activity of each neuron from the much slower dynamics of the calcium indicator. 

These problems have been traditionally treated independently in the literature. Methods of spike deconvolution have focused largely on single-pixel fluorescence data analysis. Such methods include fast nonnegative deconvolution \citep{Vogelstein10}, greedy algorithms \citep{grewe2010}, finite rate of innovation methods \citep{OSD13}, as well as particle filtering \citep{vogelstein2009spike} and MCMC methods \citep{PMP13}. While effective in the analysis of single fluorescence traces, these methods do not address the problem in a full spatiotemporal setup.

ROI identification is usually based on either of the following two observations: first, ROIs are often spatially localized, yielding methods for ROI selection based on local correlations of neighboring pixels \citep{SH10} or dictionary learning \citep{donuts}. While these methods can yield compact localized ROI estimates they do not exploit the spatiotemporal data optimally and their performance can deteriorate in the case of significant spatial overlap. A different approach stems from the observation that the spatiotemporal activity can be expressed as a product of two matrices: a spatial matrix that encodes the location of each neuron in the optical field, and a temporal matrix that characterizes the calcium concentration evolution of each neuron. Based on this observation several methods have been proposed based on independent component analysis \citep{mukamel2009}, multilevel sparse matrix factorization \citep{AH13}, and constrained nonnegative matrix factorization \citep{MMM14}. These methods can deal more effectively with overlapping sources but again do not explicitly model the calcium indicator dynamics and do not necessarily provide compact ROI estimates.

In this paper we approach all three problems simultaneously, by proposing a constrained matrix factorization method that decomposes the spatiotemporal activity into spatial and temporal components that model the dynamics of the calcium indicator and preserve the local structure of each ROI. Related methods have appeared recently in the literature. In \citet{PMG13} and \citet{PP13a} the authors propose a rank-penalized approach to initialize a matrix factorization algorithm that enforces the calcium indicator dynamics and penalizes the sparsity of each component. More recently, \citet{HYV14} propose a similar structured matrix factorization approach with a spatial total variation norm penalty to promote localized and compact ROIs. While these approaches are effective in real data analysis, they require tuning of several regularization weights, a task that can be very challenging in practice.

In our method we address this problem by proposing a constrained and structured matrix factorization approach that requires no tuning of sparsity parameters. We achieve this by introducing for each pixel a hard constraint on the energy of the residual signal between the raw data and the denoised calcium signal. These hard threshold noise levels can be estimated by exploiting the autoregressive structure of the calcium indicator dynamics. The resulting matrix factorization approach enforces the dynamics of the calcium indicator and effectively sets individual sparsity penalties for both the spiking activity and the ROI size of each neuron that are optimally tuned to satisfy the estimated residual constraints.

We present algorithms for solving this matrix factorization problem that, per iteration, scale linearly both with the total number of observed pixels and the number of timesteps, and present warm start methods that increase the computational efficiency. We also show how the methods can be parallelized to a large extent, leading to a highly efficient system that can process large movies within just a few minutes. Matrix factorization methods typically solve a bi-convex problem and their performance depends on the initialization; we propose a fast and simple greedy initialization method that detects possible neuron locations using spatial filtering methods. Finally, we introduce a few additional processing steps within the matrix factorization procedure that enhance the robustness of the algorithm and make it less sensitive to poor initialization. 

We apply our method to an {\it in vivo} large scale calcium imaging dataset and demonstrate excellent performance. Finally, to demonstrate the generality of modern matrix factorization methods in the analysis of calcium imaging data, we apply a simplified version of our algorithm to dendritic imaging data and show how such methods can effectively segment complex and dense imaging datasets.

\section{Optimal constrained deconvolution for single pixel fluorescence traces}

For ease of exposition we first address the problem of spike deconvolution from a single-pixel fluorescence times series that expresses the behavior of a single neuron. The calcium dynamics $\bm{c}$ can be approximated by a stable autoregressive process of order $p$ ($\mathrm{AR}(p)$) where $p$ is a small positive integer,
\begin{equation}
	c(t) = \sum_{j=1}^p\gamma_jc(t-j) + s(t),
	\label{AR-spikes}
\end{equation}
and $s(t)$ is the spiking signal (i.e., number of spikes) that the neuron fired at the $t$-th timestep, $t = 1,\ldots,T$. The observed fluorescence is related to the calcium concentration as:
\[
	y(t) = \alpha c(t) + b + \varepsilon_t, \quad \varepsilon_t \sim \mathcal{N}(0,\sigma^2),
\]
where $\alpha$ is a nonnegative scalar, $b$ is the baseline concentration and the noise is assumed to be i.i.d. zero mean Gaussian with variance $\sigma^2$. We assume that the baseline $b$ is known and constant, e.g. it is estimated by averaging the fluorescence over a large interval with no observed spikes. We relax both of these assumptions in the spatiotemporal case. Our goal is to perform spike inference, i.e., extract the spiking vector $\bm{s}$ from the vector of observations $\bm{y}$. 

To solve this problem we need first to estimate certain parameters. These include the order of the AR process $p$, the AR coefficients $\gamma_1,\ldots,\gamma_p$, and the observation noise variance $\sigma^2$. Assuming independent and identically distributed statistics of the spiking signal, these parameters can be estimated from standard time series analysis methods. For a given order $p$, it is easy to show that the autocovariance function of $\bm{y}$, $C_y$ satisfies the following equations:
\be
	C_y(\tau) = \left\{\begin{array}{rl} \sum_{j=1}^p\gamma_jC_y(\tau-j) - \sigma^2\gamma_{\tau}, & 1 \leq \tau \leq p \\
			 \sum_{j=1}^p\gamma_jC_y (\tau-j), & \tau > p. \end{array}\right.
	\label{covy}
\ee
By plugging the sample autocovariance values into \eqref{covy} we can first estimate the AR coefficients $\gamma_1,\ldots,\gamma_p$ and then the noise variance $\sigma^2$. For the order $p$ in general we note that if the rise time of the calcium indicator is much faster than the length of each timebin then we can safely assume that $p=1$. Otherwise, the order of the AR system can be estimated with the Akaike information criterion (AIC) \citep{Aka69}. For notational simplicity we assume $p=1$ in the following. 

\paragraph{Spike inference through optimal noise constrained deconvolution:}
Solving for the spiking vector $\bm{s}$ in the domain of nonnegative integers is a computationally hard problem. Instead, by following \citet{Vogelstein10} we can relax the spike signal to take arbitrary nonnegative values and penalize the sum of the spike signal over time to avoid overfitting. For computational reasons it is also preferable to infer the calcium signal $\bm{c}$ instead of the spiking vector $\bm{s}$. By introducing $c_1$, the initial concentration at the first timestep, and $\bm{c_{\text{in}}}$ the vector of length T given by $\bm{c_{\text{in}}}=[c_1,0,\ldots,0]^{\top}$, \eqref{AR-spikes} can be expressed in matrix form as
\be
	G(\bm{c}-\bm{c_{\text{in}}}) = \bm{s},\;\; \text{with}\;\;
G = \left[\begin{array}{ccccc} 1 & 0 & \ldots & 0 \\ -\gamma & 1  & \ldots & 0  \\ \vdots & \ddots & \ddots & \vdots \\ 0 & \ldots & -\gamma & 1\end{array}\right]. 
	\label{spike-dif}
\ee
Based on the estimate of the noise variance we can introduce a hard constraint on the energy of the residual signal to derive the following parameter-free convex program for estimating the calcium concentration up to a scaling constant:
\be
\begin{aligned}
	& \underset{\bm{c},c_1}{\text{minimize}} \quad \bm{1}_T^{\top} \bm{s}, & \\
	& \text{subject to:} \quad \bm{s} \geq 0, & \bm{s}=G(\bm{c}-\bm{c_{\text{in}}}), \quad c_1 \geq 0, 	\quad \|\bm{y} - \bm{c} - b\bm{1}_T\| \leq \sigma\sqrt{T},
\end{aligned}	
\tag{P-1$d$}
\label{opt-foopsi}
\ee
with $||.||$ denoting the 2-norm.
The inclusion of the residual as a hard constraint and not as a penalty term in the objective function (as in \citet{Vogelstein10}) allows for a parameter-free fast non-negative deconvolution approach. After solving \eqref{opt-foopsi} we can use \eqref{spike-dif} to obtain the relaxed spiking signal. Program \eqref{opt-foopsi} can be solved efficiently with a variety of methods, such as dual ascent (using the FOOPSI algorithm from \citet{Vogelstein10} in the primal step) or conic programming (using, e.g., the cvx optimization package \citep{CVX}). These methods scale linearly with the total number of timesteps $T$. Alternatively, we can also solve \eqref{opt-foopsi} directly in the spike domain using a nonnegative LARS algorithm \citep{LARS}. This LARS approach is particularly efficient when the spiking signal is expected to be very sparse, so the path following algorithm stops only after a few steps. We present details of these different approaches in the supplement.

\section{Spatiotemporal spike inference and component demixing through constrained matrix factorization}

Now we turn to the full spatiotemporal case.
At every timestep a field of view is observed, for a total number of $T$ timesteps. This field (either two- or three-dimensional) has a total number of $d$ pixels and can be vectorized in a single column vector. Thus all the observations can be described by a $d \times T$ matrix $Y$. Now assume that the field contains a total number of (possibly overlapping) $K$ neurons, where $K$ is assumed known for now. For each neuron $i$ the ``calcium activity" $\bm{c}_i$ can be described again with simple autoregressive dynamics (assumed first-order just for notational simplicity),
\begin{equation}
	c_i(t) = \gamma c_i(t-1) + s_i(t),
	\label{spikes}
\end{equation}
where $s_i(t)$ is the number of spikes that neuron $i$ fired at the $t$-th timestep, $t = 1,\ldots,T$.  Now if $\bm{a}_i \in \mathbb{R}^d_{+}$ denotes the (nonnegative) spatial ``footprint" vector for neuron $i$, then we model the spatial calcium concentration profile at time $t$ as 
\begin{equation}
	F(t) = \sum_{i=1}^K \bm{a}_ic_i(t) + B(t),
	\label{fluor}
\end{equation}
where $B(t) \in \mathbb{R}^d_{+}$ denotes the (time-varying) baseline vector all the pixels. Finally, at each timestep we observe $F(t)$ corrupted by additive Gaussian noise:
\be
	Y(t) = F(t) + \bm{\varepsilon}_t, \quad  \bm{\varepsilon}_t \sim \mathcal{N}(\bm{0},\Sigma),
	\label{obs}
\ee
where $\Sigma$ is a diagonal matrix (indicating that the noise is spatially and temporally uncorrelated). Eqs. \eqref{spikes}-\eqref{obs} can be written in matrix form as\protect\footnote{We ignore the initial values for simplicity.}
\[
	\begin{split}
		S &= CG^{\top}\\
		F &= AC + B\\
		Y & = F + E,
	\end{split}
\]
with  $S = [\bm{s}_1,\ldots, \bm{s}_K]^{\top}$, $C = [\bm{c}_1,\ldots, \bm{c}_K]^{\top}$, $A = [\bm{a}_1,\ldots,\bm{a}_K]$, $F=[F(1),F(2),\ldots,F(T)]$, $Y = [Y(1),Y(2), \ldots, Y(T)]$, $B = [B(1),B(2), \ldots, B(T)]$. In practice, we have found that the background activity matrix $B$ can often be modeled as a rank 1 matrix, $B = \bm{b}\bm{f}^{\top}$, where $\bm{b}\in\mathbb{R}_+^d$, $\bm{f}\in\mathbb{R}_+^{\top}$ are nonnegative vectors encoding the background spatial structure (typically consisting of a sum of baseline activity from the neurons of interest and densely mixed neuropil structure below the observed spatial resolution) and global (possibly time varying) intensity, respectively.\protect\footnote{Higher rank terms can also be used here if necessary.}

\subsection{Optimal matrix factorization deconvolution methods}
\label{nmf}

Assuming the number of neurons $K$ and initial estimates of $A,C$ and $\bm{b},\bm{f}$ as known, we can apply alternating matrix factorization methods to estimate the spatial components $A,\bm{b}$ given the temporal $C,\bm{f}$ and vice versa, from the fluorescence observations $Y$. We present an efficient initialization procedure in section \ref{nmf:in}.

\paragraph{Estimating $A, \bm{b}$:} Since each column of $A$ expresses the location of a neuron, we want $A$ to be sparse to promote localized spatial footprints. Given estimates of $C$ and $\bm{f}$ from the previous iteration, the spatial matrix $A$  and background $\bm{b}$ can be updated by solving the following convex program
\begin{flalign}
	 \underset{A}{\text{minimize}} & \; \|A\|_1,  \nonumber \\
	 \text{subject to:} & \; A,\bm{b} \geq 0, \quad  \|Y(i,:) - A(i,:)C - b(i)\bm{f}^{\top} \| \leq \sigma_i\sqrt{T}, \; i = 1,2,\ldots,d \label{spatial} \tag{P-S}.
\end{flalign}
where $A(i,:),Y(i,:)$ denote the $i$-th rows of $A$ and $Y$ respectively. Although the matrix $A$ is of very large size, $d\times T$, the problem \eqref{spatial} can be readily parallelized into $d$ programs for each pixel separately. Each of these problems can be solved either by using the non-negative LARS algorithm or by using a dual ascent method. LARS is preferred here, since the number of neurons that overlap in a given pixel (and therefore the dimension of the resulting LARS problem) is in general very small. 

When the fluorescence from each neuron is highly localized near the soma, the process of estimating $A$ at the $k$-th iteration can be further sped up by using the previous estimate $A^{k-1}$ as follows. After computing $A^{k-1}$, we can approximate the center and the size of each neuron (expressed by a column of $A$), by computing its center of mass and the variance around this center. Then the support of this cell's ROI can be computed as an ellipse centered at the center of mass and rotated along the two principal components that capture most of the variance of the mass of the neuron. When estimating the $i$-th row of $A^k$, we can restrict our search to the neurons (columns of $A$) whose ROIs include the pixel $i$.  This sparsens $A^k$ significantly and makes the dimensionality of each LARS subproblem much smaller, leading to a highly efficient and parallelizable update.

Note that we have not yet incorporated any prior information about the detailed shape of the spatial components $A(i,:)$, which enabled the highly parallel approach described above.  However, in many cases it is natural to assume that $A(i,:)$ is connected, or smooth in a suitable sense.  Empirically we have found it helpful to include a mild post-processing step at each iteration, using standard non-linear image filtering techniques, such as median filtering or morphological opening, which are effective in removing isolated pixels that appear as active. The removed pixels can then be absorbed by the background component.

\paragraph{Estimating $C,\bm{f}$:} For the temporal components we want to introduce a sparsity penalty to the spiking signal of each neuron to prevent overfitting. We can again use our estimates of the noise variance as hard constraints and derive a parameter-free convex program:
\begin{flalign}
	\underset{\bm{c}_1,\ldots,\bm{c}_K,\bm{f}}{\text{minimize}} &\;\; \sum_{j=1}^K\bm{1}^{\top}G\bm{c}_j, \nonumber \\
	\text{subject to:} &\;\; G\bm{c}_j \geq 0, \; j = 1,2,\ldots,K   \label{temporal}  \tag{P-T} \\
	& \; \|Y(i,:) - A(i,:)C - b(i)\bm{f}^{\top} \| \leq \sigma_i\sqrt{T}, \; i = 1,2,\ldots,d \nonumber. 
\end{flalign}
Since the constraints $G\bm{c}_i \geq 0$ couple the entries within each row of $C$, and the residual constraints within each column, the program \eqref{temporal} cannot be readily parallelized. Moreover, the large number of constraints and the potentially large number of neurons $K$ make the direct solution of \eqref{temporal} computationally expensive. To overcome this we employ a block-coordinate descent approach where we sequentially update the temporal component $\bm{c}_j$ of each neuron. If we denote by $A_{:\backslash j}$ ($C_{\backslash j :}$) the matrix $A$ ($C$) with its $j$-th column (row) removed, and $F_{\backslash j} = A_{:\backslash j}C_{\backslash j :} 
+ \bm{bf}^{\top}$, then each $\bm{c}_j$ can be found by solving
\begin{flalign}
	\underset{\bm{c}_j}{\text{minimize}} &\;\; \bm{1}^{\top}G\bm{c}_j, \nonumber \\
	\text{subject to:} &\;\; G\bm{c}_j \geq 0, \quad \|Y(i,:) - F_{\backslash j}(i,:) - a_j(i)\bm{c}_j^{\top} \| \leq \sigma_i\sqrt{T}, \; i = 1,2,\ldots,d  \label{temporal_coord}. 
\end{flalign}
To solve \eqref{temporal_coord} we employ a dual ascent method where we introduce a set of Lagrange multipliers $\bm{\lambda}_j\in\mathbb{R}^d$ and consider the function 
\[
	\mathcal{L}(\bm{c}_j,\bm{\lambda}_j) = \left\{\begin{array}{rr}  \bm{1}^{\top}G\bm{c}_j + \sum_{i=1}^d\lambda_j(i)(\|Y(i,:) - F_{\backslash j}(i,:)  - a_j(i)\bm{c}_j^{\top} \|^2 - \sigma_i^2T), & G\bm{c}_j \geq 0 \\ \infty, & \text{otherwise} \end{array}\right.
\]
For a given value of  $\bm{\lambda}_j$, $\mathcal{L}(\bm{c}_j,\bm{\lambda}_j)$ can be easily minimized with respect to $\bm{c}_j$ in $O(T)$ time using the FOOPSI log-barrier interior point method. Then the vector $\bm{\lambda}_j$ can be updated as
\[
	\bm{\lambda}_j^k = \bm{\lambda}_j^{k-1} - \alpha_k\nabla_{\bm{\lambda}_j}\mathcal{L}(\bm{c}^k_j,\bm{\lambda}_j).
\]
Note that at iteration $k$, $\bm{\lambda}_j$ can be warm started with the value $\bm{\lambda}_j^{k-1}$. Moreover, to increase speed we can consider only a small number of the hard constraints, e.g., corresponding to the pixels where the spatial component $\bm{a}_j$ has the highest values. After we update all the temporal components we also update the background activity vector $\bm{f}$ by solving a simple nonnegative least squares program.  (Note that we do not enforce a sparsity constraint on either $\bm{f}$ or $\bm{b}$.)

This block-coordinate descent approach provides a tractable update for the estimates of the temporal component $C$. Although this method is in general not parallel it can be parallelized to a large extent by exploiting the fact that the different ROIs as specified through the columns of the matrix $A$ do not overlap significantly. At every iteration we can form a graph where each vertex corresponds to a neuron and   two neurons are connected with an edge if their spatial filters overlap. Then we can update the temporal components as follows: First, we find the minimum set of vertices such that when removed, the graph is split into a set of disjoint subgraphs. This problem is known as the minimum vertex cover problem in graph theory. It is known to be NP-complete, although for sparse graphs certain approximation algorithms perform well in practice \citep{Vaz01}. After we find this set, we update the temporal components of these neurons, and then we repeat this process in each of the subgraphs in parallel. We can repeat this temporal block-coordinate descent approach until convergence, which typically occurs only after a few iterations. 

\paragraph{Merging existing ROIs:}

Depending on the initialization procedure, a neuron can sometimes be initially split into two or more different ROIs, that subsequently need to be merged. To detect ROIs that need to be merged, we again construct a graph where each vertex corresponds to a neuron and   two neurons are connected with an edge if their ROIs overlap. For this graph we detect all the maximal cliques, i.e., the cliques of the graph that are not part of larger cliques. This is again an NP-complete problem which can nevertheless be solved efficiently for large sparse graphs \citep{ELS10}. Now for each of these maximal cliques we compute the correlation matrix of the temporal components of the corresponding nodes. We find the largest principal submatrix where all the correlation coefficients are above a certain threshold, and merge the corresponding ROIs. A similar merging procedure is also performed when a component is significantly correlated with the background activity, in which case the component is absorbed into the background.

\paragraph{Removing ROIs:}

After each iteration we can remove any ROIs that do not contribute significantly to the overall spatiotemporal activity, by discarding ROIs that are spatially nearly empty and/or do not contribute any spiking activity.  After the last iteration, we normalize each spatial component to have unit energy, and then sort the components based on the maxima of their temporal factors. Empirically we find that this approach effectively orders the neurons, with the first components in the list having strong but sparse temporal activity, making it easy for the user to set a cutoff $\hat K$ above which the obtained activity is retained, and below which the components are discarded.

\subsection{Greedy initialization of the matrix factorization approach}
\label{nmf:in}

The matrix factorization approach presented in section \ref{nmf} allows us to effectively separate the different neurons and deconvolve their spikes in a computationally tractable way. However, this approach is bi-convex and can converge to a local maximum point that depends on the specific initialization of the matrix factorization procedure. We designed a custom greedy method for approaching this problem in a computationally efficient way. At every iteration the spatiotemporal data matrix is spatially filtered with a Gaussian kernel of width similar to the size of the neuron. The algorithm finds the location where this filtering procedure explains the maximum variance and draws a square ROI  of size roughly twice the size of an average neuron. Within this ROI a rank 1 nonnegative matrix factorization (initialized with the rank-1 SVD of this small patch) is performed to initialize the spatial and temporal components, and the product of these components is then subtracted from the observed data. This procedure is repeated until a user specified number of neurons is located. Then, the resulting residual signal is used to initialize the background component $B = \bm{bf}^{\top}$  using rank-1 nonnegative matrix factorization. As discussed above, the total number of neurons is unknown and in practice may be difficult to estimate automatically. A natural strategy is to begin with an overly large value of $K$, subsequently merging or removing components after each constrained matrix factorization step, or the weakest neurons (defined by the ordering described above) can be discarded in the final step of the algorithm, with very modest user input (a simple choice of the cutoff value $\hat K$).  A full description of this initialization procedure is given in the supplement. Our full procedure for spike inference and component demixing is schematically represented as Algorithm 1.

\begin{algorithm}[t]
	\caption{Constrained matrix factorization for spatiotemporal spike inference}
	\begin{algorithmic}[1]
	\State Initialize $A$, $C$, $\bm{b}$, $\bm{f}$ using the greedy initialization approach.
	\Repeat
	\State Compute center of mass for each spatial component and define search region for each pixel.
	\State Update spatial components $A$ and background component $\bm{b}$ by solving \eqref{spatial}.
	\State Perform median filtering for each component. Absorb discarded pixels into the background.
	\State Update temporal components $C$ and background activity $\bm{f}$ \eqref{temporal} using coordinate descent.
	\State Merge overlapping spatial components with overly highly correlated temporal components.
	\State Remove overly weak components.
	\Until{convergence}
	\State{Order components and let user choose cutoff value $\hat{K}$.}
	\end{algorithmic}
\end{algorithm}

\section{Results}
\begin{figure}[t!]
	\centering
	\includegraphics[trim = 6mm 16mm 34mm 6mm, clip, width=\textwidth]{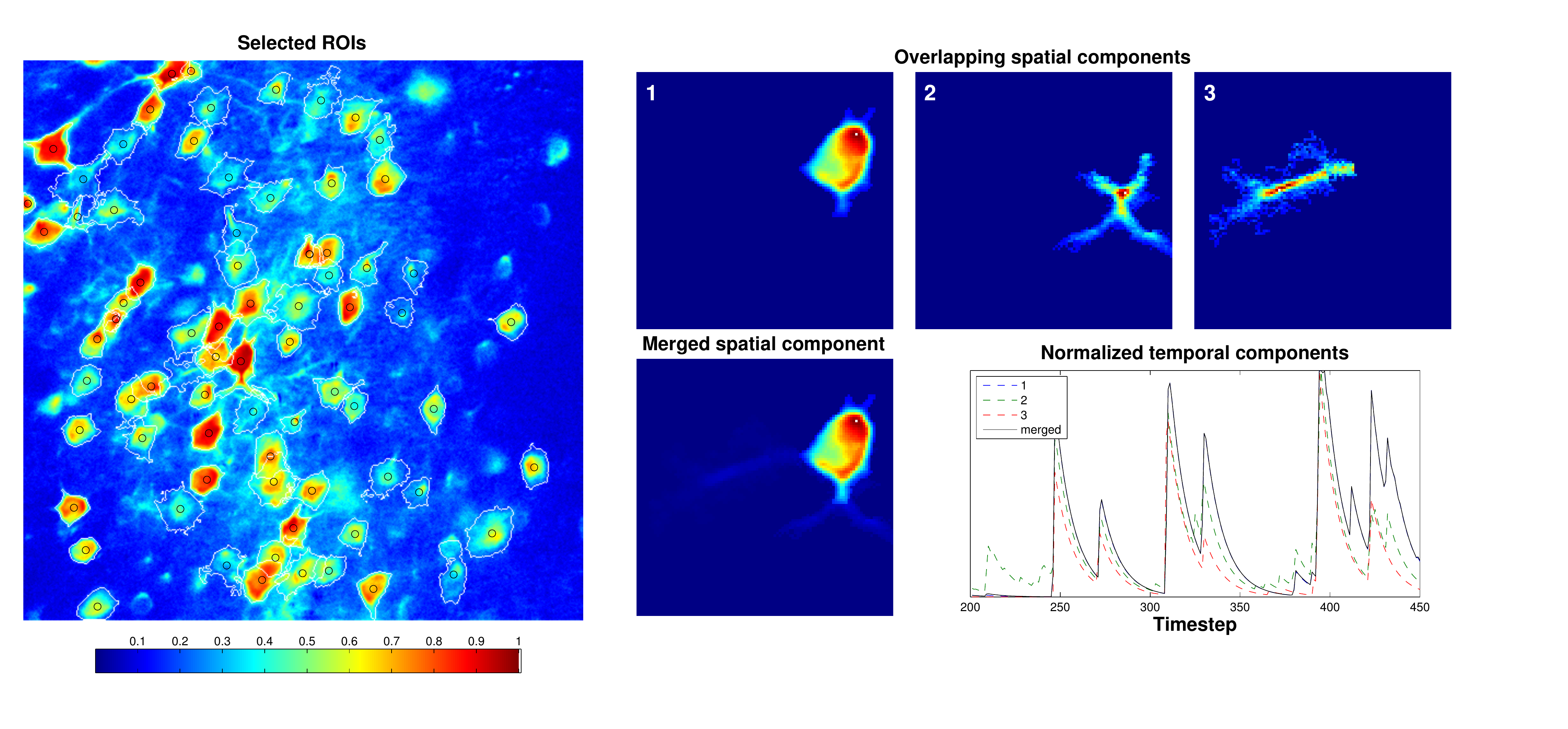}
	\caption{Application to mouse V1 {\it in vivo} data. Left: Inferred ROIs superimposed on the correlation image of the raw data. The white contours show the boundary of each inferred cell shape; black circles denote the corresponding centers of mass. The supplementary video provides a more complete depiction of the results. Right: Depiction of the merging operation: Upper panels: Three overlapping spatial ROIs with highly correlated temporal components. Lower left: Merged spatial component. Lower right: Estimated temporal components of individual (dashed) and merged (solid) components.}
	\label{kira1}
\end{figure}

\paragraph{Application to population imaging data:} We begin by applying our methods to {\it in vivo} mouse V1 spontaneous activity data. The results are shown in Fig.~\ref{kira1}. The algorithm was initialized with the procedure described above with 85 components. During the factorization iterations 10 components were eliminated due to merging operations or negligible total contribution. The contours of the final 75 ROIs are depicted in the left panel of Fig.~\ref{kira1}, superimposed on the ``correlation image" \citep{SH10} of the raw data. The correlation image for each pixel is computed by averaging the correlation coefficients (taken over time) of each pixel with its 8 immediate neighbors. Localized regions in the correlated image with high intensity correspond to strongly active cells, whereas localized regions with lower intensity correspond to neurons with lower intensity or other non-stationary processes. The algorithm efficiently identifies neurons with very few visually-apparent false positives. The results are viewed best in the video included in the supplementary material. The remaining panels display an example of the merging procedure. The neuron depicted in the lower panel is initially split across three components (upper panels; one for the soma and two for identified dendrites). Since the temporal activity of these components is highly correlated they are merged into a single cell.  

\begin{figure}[t!]
	\centering
	\includegraphics[trim = 2mm 14mm 32mm 4.5mm, clip, width=\textwidth]{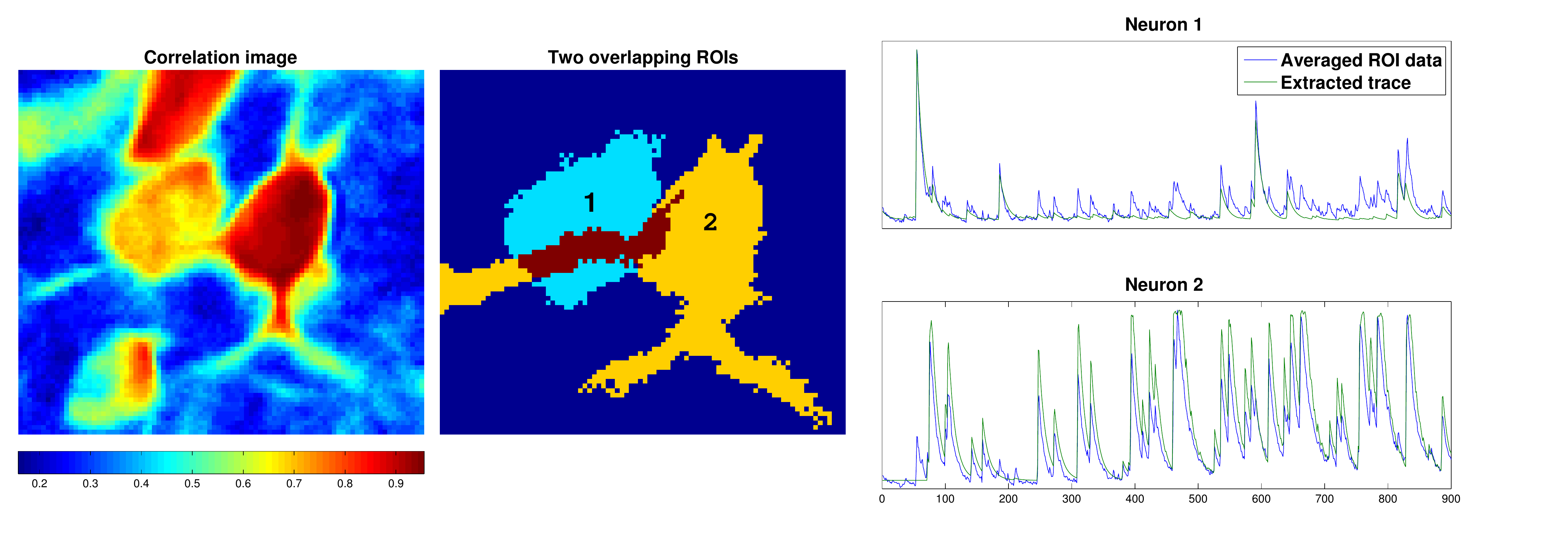}
	\caption{Importance of the demixing procedure. Left: Correlation image zoomed into two overlapping neurons. The exact spatial masks cannot be readily resolved. Middle: Spatial masks for two identified overlapping ROIs after demixing (the brown region corresponds to the overlapping pixels). Right: Inferred calcium traces of the two neurons obtained by averaging over the spatial masks (blue traces) and by applying the proposed method (green traces). Simple averaging over the ROI of neuron 1 can be misleading if demixing is not performed.  
	}
	\label{kira2}
\end{figure}

Fig.~\ref{kira2} highlights the importance of the demixing procedure. The two identified neurons in the left panel overlap over the region displayed in brown. When analyzing neuron 1, averaging over all of its spatial mask (without first excluding the activity of neuron 2) introduces false spikes (as depicted in the blue trace of the upper right panel) since the activity of neuron 2 is much stronger and influences the analysis even when a relatively small overlap is present. 


\paragraph{Application to dendritic imaging data:}
A key advantage of the proposed structured matrix factorization framework is that we can apply similar methods to dendritic imaging data where the imaging focuses on the dendrites of multiple neurons and not on the cell bodies. In this case each spatial component corresponds to a set of dendritic branches from a given neuron and the temporal component to the synchronous activity of these branches. The goal is to segment these movies and disentangle the various dendritic branches. These images exhibit certain qualitative differences compared to somatic imaging. Each spatial component is again sparse but is no longer spatially localized since dendritic branches can stretch significantly along the observed imaging plane. As a result, the degree of overlap between the different branches is significantly higher, making even rough interpretation by eye a challenging task; the correlation image in this setting provides very little useful segmentation information. Moreover, the bound calcium dynamics no longer follow somatic calcium indicator dynamics since they are affected by the highly nonlinear processing that takes place in the dendrites. Dropping the temporal dynamics and spatial localization constraints from our problem, we obtain a simpler sparse nonnegative matrix factorization problem which can still be solved efficiently using the methods described above.

\begin{figure}[!t]
	\centering
	\includegraphics[trim = 5mm 13mm 46mm 5mm, clip, width=\textwidth]{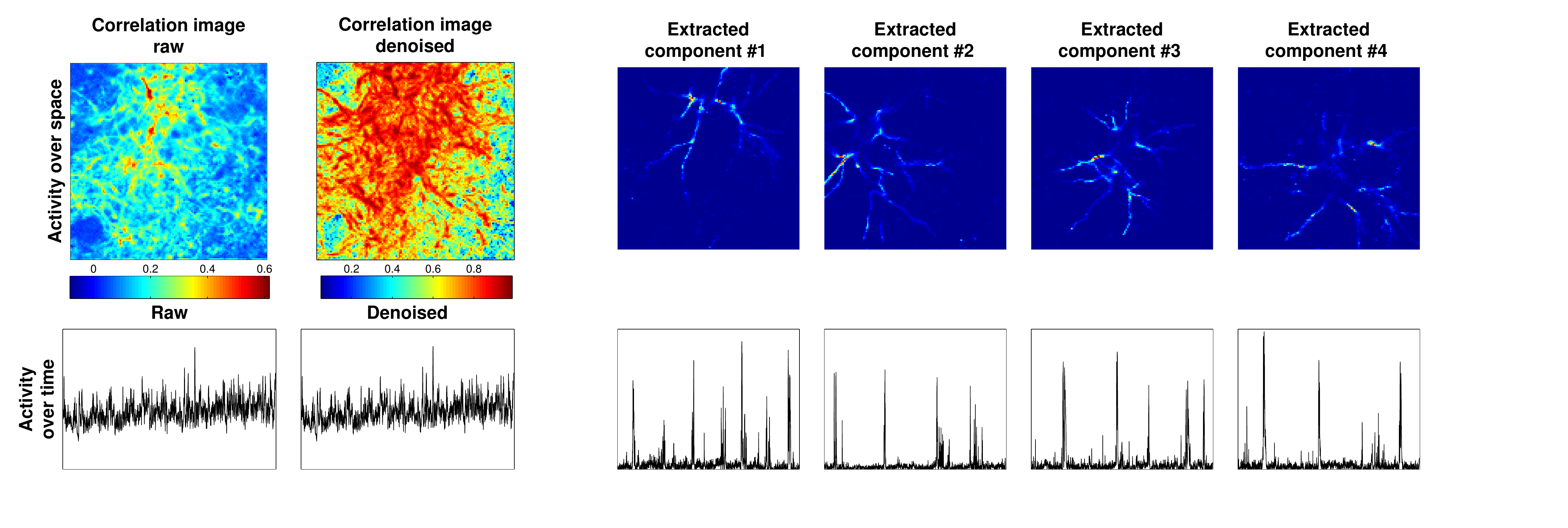}
	\caption{Application to {\it in vivo} dendritic imaging data from rodent barrel cortex. Top: Correlation images of the raw and denoised data, and four extracted components. Bottom: Spatially averaged activity over time of the raw and denoised data and the temporal traces of the 4 extracted components. The proposed method can segment the dense dendritic imaging data and reveal a rich underlying sparse structure. The supplementary video provides a more complete depiction of the results.}
	\label{clay}
\end{figure}

We applied this approach to {\it in vivo} dendritic imaging data taken from the apical dendrites of layer-5 pyramidal neurons in the rodent barrel cortex and show the results in Fig.~\ref{clay}. The raw data is typically dense both in time and space as can be seen from the low intensity of the raw data correlation image (upper left panel). We initialize using a large number of components (in this case 50), and then select the components that correspond to actual dendritic structure using the sorting procedure described above.  The resulting denoised movie displays much stronger local correlations (Fig.~\ref{clay} second column), with the top 20-30 inferred components corresponding to localized dendritic structures that are sparsely active at specific points in time. Four of these components are shown in Fig.~\ref{clay}. A more informative movie of the results as well as a depiction of the sorted ROIs and temporal components can be found in the supplementary material.  

To conclude, these results demonstrate that the proposed matrix factorization methods provide powerful tools for analyzing large scale calcium imaging datasets and extracting informative spatiotemporal components. 

\appendix

\section{Algorithms for solving the single pixel constrained deconvolution problem}

We briefly discuss the three different approaches that can be used to solve program \eqref{opt-foopsi}, which we repeat here in an equivalent form for completeness:
\be
\begin{aligned}
	& \underset{\bm{c},c_1}{\text{minimize}} &\bm{1}_T^TG(\bm{c}-\bm{c_{\text{in}}}), \\
	& \text{subject to:} & G(\bm{c}-\bm{c_{\text{in}}}) \geq 0, \quad c_1 \geq 0 \\
	& & \|\bm{y} - \bm{c} - b\bm{1}_T\| \leq \sigma\sqrt{T}.
\end{aligned}	
\tag{P-1$d$}
\label{opt-foopsi}
\ee

\paragraph{Dual ascent methods:}

We introduce Lagrange multipliers for the constraints and define as $\bm{c}^{\lambda},c_1^{\lambda}$ as the solution to the following program
\be
\begin{aligned}
	& \underset{\bm{c},c_1}{\text{minimize}} & \mathcal{L}(\bm{c},\bm{c_{\text{in}}},\lambda) =  \bm{1}^TG(\bm{c}-\bm{c_{\text{in}}}) + \lambda(\|\bm{y}-\bm{c}-b\bm{1}_T\|^2 - \sigma^2T), \\
	& \text{subject to:} & G(\bm{c}-\bm{c_{\text{in}}}) \geq 0 \quad c_1 \geq 0. 
\end{aligned}	
\label{dual}
\ee
The problem \eqref{dual} can be readily solved in $O(T)$ time with the interior point method of \citet{Vogelstein10}. After solving \eqref{dual}, the Lagrange multiplier can be updated as
\be
	\lambda_k = \lambda_{k-1} - a_{k}\nabla_{\lambda}\mathcal{L}(\bm{c}^{\lambda_{k-1}},\bm{c_{\text{in}}}^{\lambda_{k-1}},\lambda) = \lambda_{k-1} - a_{k}(\|\bm{y}-\bm{c}-b\bm{1}_T\|^2 - \sigma^2T),
\ee
where $a_{k}$ is an appropriate step size, determined e.g. by line search.

\paragraph{Conic programming:} The program of \eqref{opt-foopsi} can also be solved with standard interior point methods for conic programming. Due to the simplicity of the residual and non-negativity constraints the solution can be efficiently computed in $O(T)$ using standard computational methods, e.g. the CVX computational package \citep{CVX}.

\paragraph{Nonnegative LARS:} The problem can also be solved directly in the spike domain using a nonnegative LARS algorithm \citep{LARS}. More specifically we consider the modified problem in the spike domain as follows
\be
\begin{aligned}
	& \underset{\bm{s}}{\text{minimize}} &\frac{1}{2\sigma^2} \|\bm{y} - G^{-1}\bm{s}-b\bm{1}_T\|^2 + \lambda\bm{1}^T\bm{s}, \\
	& \text{subject to:} & \bm{s} \geq 0 \\
	& & \|\bm{y} - G^{-1}\bm{s}-b\bm{1}_T\| \leq \sigma\sqrt{T}.
\end{aligned}	
\label{LARS-foopsi}
\ee
The solution path is computed in the standard piecewise linear way starting from $\lambda_0 = \infty$. As $\lambda$ decreases, more spikes are added in the solution reducing the energy of the residual signal. The path algorithm is stopped when the produced solution satisfies the residual constraint with equality. Let $\lambda_{k-1},\lambda_k$ be the values of $\lambda$ at the $(k-1)$-th and $k$-th step of the algorithm respectively, and $\bm{s}^{k-1}$, $\bm{s}^{k}$ the corresponding solutions. $k$ is chosen such that 
\[ \|\bm{y} - G^{-1}\bm{s}^k -b\bm{1}_T\| \leq \sigma\sqrt{T} \leq \|\bm{y} - G^{-1}\bm{s}^{k-1}-b\bm{1}_T\|. \]  
Between the $(k-1)$-th and $k$-th steps the solution changes according to the direction
\[ \bm{s}_{\text{dir}} = - (\bm{s}^k - \bm{s}^{k-1})/(\lambda^k - \lambda^{k-1}),\]
and the solution $\bm{s}^{\ast}$ can be found by finding the solution $\lambda^{\ast}$ of the quadratic equation
\[ \|\bm{y} - \lambda G^{-1}(\bm{s}^{k-1} + \bm{s}_{\text{dir}})-b\bm{1}_T\|^2 = \sigma^2T, \]
and setting
\be
	\bm{s}^{\ast} = \bm{s}^{k-1} + (\lambda^{k-1}-\lambda^{\ast})\bm{s}_{\text{dir}}.
\ee
Note that this approach does not differentiate between the initial concentration $\bm{c_{\text{in}}}$ and the spiking signal $\bm{s}$. However this difference does not affect the solution significantly beyond the first timestep. 

The LARS approach is particularly efficient when the spiking signal is expected to be very sparse so the algorithm stops only after a few steps.  For more dense spiking, the LARS algorithm can require more steps to achieve the constraints and thus it can be inefficient. However, the LARS algorithm can produce a solution even when the problem \eqref{opt-foopsi} is infeasible. In this case, the LARS algorithm computes the full path and produces a (dense) solution for $\lambda=0$ that satisfies the nonnegativity constraints.

\section{Captions for supplementary videos and supplementary figure }

\paragraph{Video S1: \url{http://www.stat.columbia.edu/~eftychios/movies/Kira-vid2.mp4} \\} 
Application to GCaMP6s-expressing neurons in cortical layer 2/3 of adult mouse V1. Expression was achieved via viral injection of AAV1-hsyn-GCaMP6s into C57Bl/6 mice, three weeks prior to imaging. Frame-scanned, two-photon imaging (950nm excitation wavelength, 535/50 emission filter) was carried out at 3Hz using a 25x (1.05 N.A.) objective. No motion correction was done on the data.
Top row: Left: Raw data, Middle: Denoised data without synchronized background activity, Right: Residual signal at 4$\times$ finer scale (the synchronized background activity is not included in the residual). Bottom left: Background synchronized activity. For the rest of the panels, 4 representative extracted spatiotemporal components (top) and the corresponding patches of the raw data. The algorithm successfully denoises the signal and demixes the overlapping neurons.

\paragraph{Video S2: \url{http://www.stat.columbia.edu/~eftychios/movies/clay-denp-color.mp4} \\} 
Application to calcium signals from apical dendrites of cortical Layer 5 pyramidal neurons were obtained by injecting AAV2/9-hSyn-FLEX-GCaMP6f (UPENN vector core) into the barrel cortex of Rbp4:Cre BAC transgenic mice (GENSAT). Two-photon imaging was performed at 4Hz with a 16x, 0.8NA lens (Nikon) at 940nm while mice performed a whisker-based object detection task. Resulting TIF stacks were motion corrected with a dynamic programming algorithm presented in \citet{kaifosh2013}.
 Top row: Left: Raw data, Middle: Denoised data with the background and noisy components removed. Right: Residual signal at 2$\times$ finer scale. Bottom panels: 7 of the spatiotemporal extracted components plus the background synchronized activity (lower right panel). The video contains only the frames where at least one of the displayed components is significantly active. The algorithm extracts rich and structured spatiotemporal components that are not visible by plain observation of the raw data.

\begin{figure}
	\centering
	\includegraphics[trim = 70mm 16mm 52mm 9mm, clip, width=\textwidth]{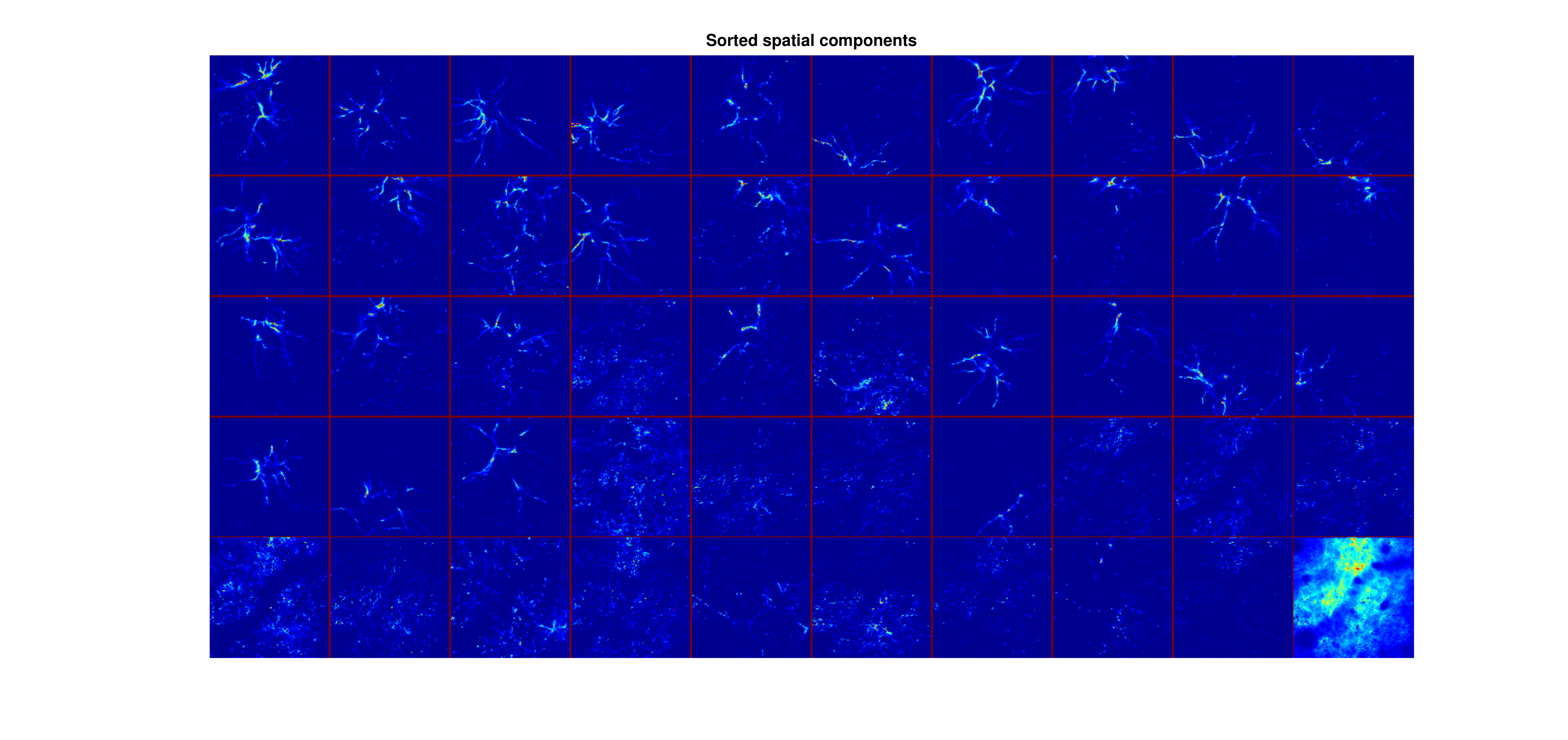} \\
	\includegraphics[trim = 50mm 16mm 42mm 9mm, clip, width=\textwidth]{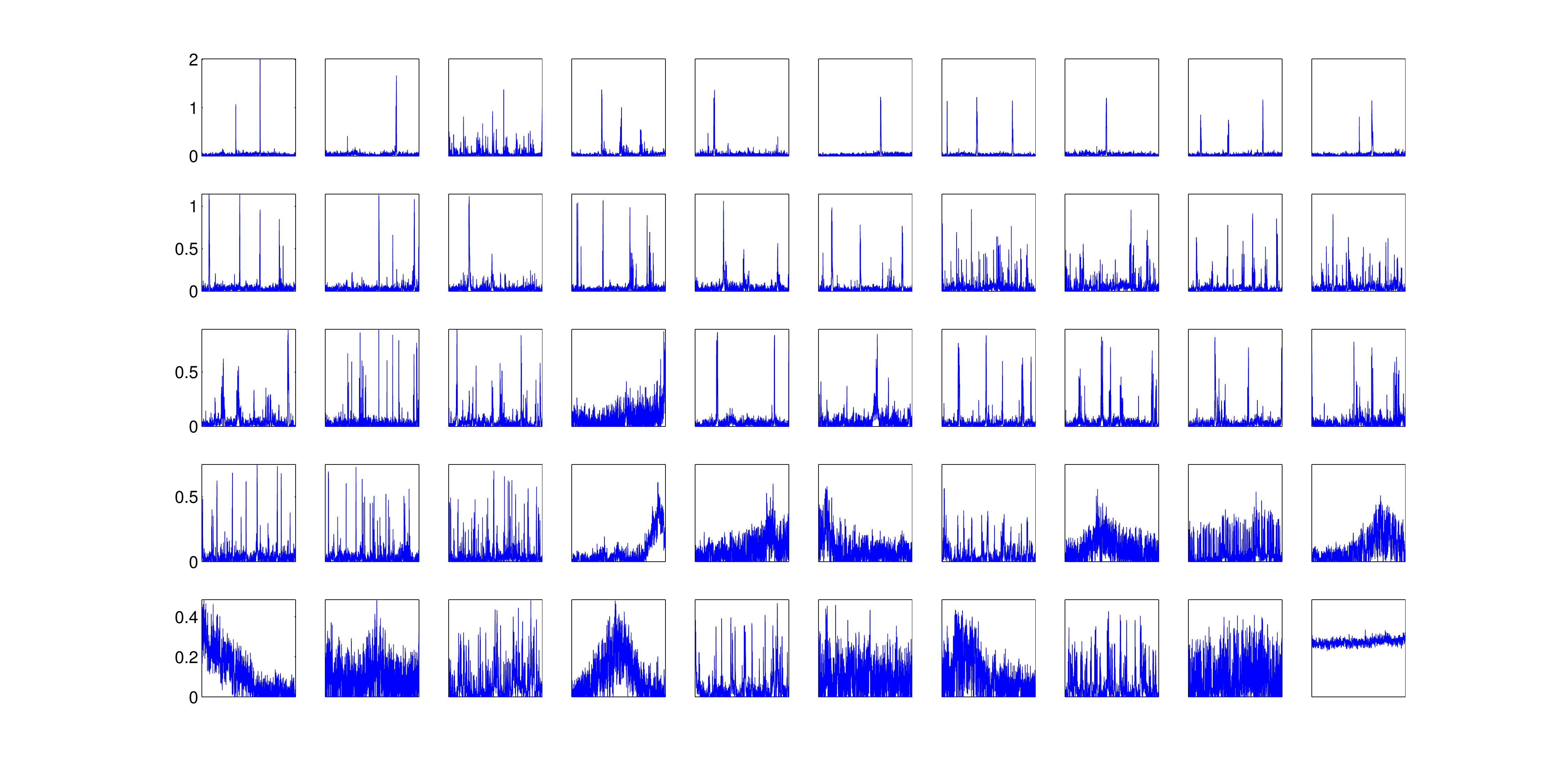}
	\caption{Sorting of the spatial components (top) based on the maximum of their temporal activity (bottom). The structured dendritic extracted components are typically sorted before the noisy components because of their activity is sparse in time but with high magnitude, indicating actual dendritic activity.}
\end{figure}

\newpage
\section{Algorithmic description for the greedy initialization procedure}

\begin{algorithm}
\caption{Greedy neuron identification}
\label{alg:greedy}

\begin{algorithmic}[t1]
\Require Data $Y \in \mathbb{R}^{d \times T}$; number of neurons needed $K$; standard deviation of the 2-D Gaussian kernel used to scan $\tau = (\tau_x, \tau_y)$; window size $w = (w_x, w_y)$.
\Procedure{GreedyNeuronId}{$Y$, $K$, $\tau$, $w$}
\State $R = Y$;
\State Define Gaussian blur matrix $D \in \mathbb{R}^{d \times d}$, where column $i$ is a (vectorized) truncated 2-D Gaussian kernel centered at pixel $i$ with variance $(\tau_x^2, \tau_y^2)$, supported in a $w_x \times w_y$ window centered at $i$ $(1 \leq i \leq d)$;
\For{$i=1:d$}
	\State Subtract and store median value for each pixel, $m(i) = \mathrm{Median}(Y(i,:))$.
\EndFor
\For{$k = 1:K$}
	\State Calculate variance explained by each kernel, $\rho = D^T R$, $v_i = \sum_{t = 1}^T \rho_{it}$;
	\State Identify the center of neuron $k$, $i_k = \arg \max_i v_i$
	\State Define $S_k$ to be the set of all pixels lie in the $w_x \times w_y$ window centered at $i_k$, solve
	\begin{equation}
	\label{equ:r1pca}
	\begin{aligned}
	& \underset{\bm{a}_k \in \mathbb{R}^{d}, \bm{c}_k \in \mathbb{R}^{T}}{\text{minimize}}
	& & \| R - \bm{a}_k \bm{c}_k^T \|^2\\
	& \text{subject to:}
	& & a_{k}(i) \geq 0, i \in S_k\\
	&
	& & a_{k}(i) = 0, i \notin S_k. 
	\end{aligned}
	\end{equation}
	\State $R \leftarrow R - \bm{a}_k \bm{c}_k^T$;
\EndFor
	\State $R \leftarrow R + \bm{m} \bm{1}_T^T$. Add median values back to the residual and solve
	\begin{equation}
	\begin{aligned}
	& \underset{\bm{b} \in \mathbb{R}^{d}, \bm{f} \in \mathbb{R}^{T}}{\text{minimize}}
	& & \| R - \bm{b}\bm{f}^T \|^2\\
	& \text{subject to}
	& & b(i) \geq 0,\; i=1,\ldots, d\\
	&
	& & f(t) \geq 0,\; t =1, \ldots, T. \\
	\end{aligned}
	\end{equation}
\State {\bf return} $A = [\bm{a}_1, ..., \bm{a}_K]$, $C = [\bm{c}_1, ..., \bm{c}_K]^T$, $\bm{b}$, $\bm{f}$.
\EndProcedure
\end{algorithmic}
\end{algorithm}
At the beginning we center the data at each pixel around zero by subtracting the median over time. At each iteration, we use a (truncated) Gaussian kernel of size similar to a neuron to scan the residual and identify the location where the kernel explains the most variance over time. Then a rank-1 matrix factorization extracts the spatial component that is localized around the identified location. After neuron $k$ has been identified, the inferred signal is subtracted to update the residual for the next neuron. When all $K$ neurons have been identified, the median is added back to the residual signal and the background activity is estimated using by solving a rank-1 nonnegative matrix factorization problem.

Note that since the spatial component is localized, at each step only a small portion of the residual is updated and therefore only a small portion of the explained variance $v_i$ needs updating. Also note that the matrix factorization step (equation (\ref{equ:r1pca}) in algorithm \ref{alg:greedy}) can be efficiently done by alternating between optimizing $\bm{a}_k$ and $\bm{c}_k$, since the Gaussian kernel scan gives a reasonable initialization. Usually 5 iterations are enough for convergence. solution of equation (\ref{equ:r1pca}) is non-identifiable by a scalar multiplication, and we can simply identify the result by constraining spatial components to have unit norm.

\bibliographystyle{chicagoa} 

\end{document}